\documentclass[
reprint,
amsmath,
pra,
superscriptaddress,
noeprint, longbibliography,
]{revtex4-2}
\usepackage{siunitx}
\usepackage{graphicx}
\usepackage{dcolumn}
\usepackage{braket}
\usepackage{xfrac}
\usepackage{hyperref}
\usepackage[usenames,dvipsnames]{color}
\usepackage{tikz-cd}
\usepackage{float}
\usepackage[nameinlink,capitalise]{cleveref}
\usepackage{mathtools}
\usepackage{amsmath}
\usepackage{bm}
\usepackage{color}
\DeclareSIUnit\gauss{G}
\DeclareSIUnit{\au}{{a.u.}}
\hypersetup{
colorlinks=true,
linkcolor=blue,
citecolor=blue,
filecolor=green,
urlcolor=blue,
}

\begin{document}
\title{Full-dimensional quantum scattering calculations on ultracold atom-molecule collisions in magnetic fields:  The role of molecular vibrations}

\author{Masato Morita}
\affiliation{Department of Physics, University of Nevada, Reno, NV, 89557, USA}
\author{Jacek K\l os}
\affiliation{Department of Chemistry and Biochemistry, University of Maryland College Park, College Park, Maryland, 20742, USA}
\altaffiliation{Current affiliation: Department of Physics, Joint Quantum Institute, University of Maryland College Park, College Park, Maryland, 20742, USA}
\altaffiliation{Department of Physics, Temple University, Philadelphia, Pennsylvania 19122, USA}
\author{Timur V. Tscherbul}
\affiliation{Department of Physics, University of Nevada, Reno, NV, 89557, USA}

\begin{abstract}
Rigorous quantum scattering calculations on ultracold molecular collisions in external fields present an outstanding computational problem due to strongly anisotropic atom-molecule interactions that depend on the relative orientation of the collision partners, as well as on their vibrational degrees of freedom.
Here, we present the first numerically exact three-dimensional quantum scattering calculations on strongly anisotropic atom-molecule (Li~+~CaH) collisions in an external magnetic field based on the parity-adapted total angular momentum representation and a new three-dimensional potential energy surface (PES) for the triplet Li-CaH collision complex using the unrestricted coupled cluster method with single, double and perturbative triple excitations [UCCSD(T)] and a large quadruple-zeta type basis set.
We find that  while the full three-dimensional treatment  is necessary for the accurate description of  Li($M_S=1/2$)~+~CaH($v=0,N=0,M_S=1/2$) collisions as a function of magnetic field, the magnetic resonance density and statistical properties of spin-polarized  atom-molecule collisions are not strongly affected by vibrational degrees of freedom, justifying the rigid-rotor approximation used in previous calculations. We observe rapid,  field-insensitive vibrational quenching in ultracold  Li($M_S=1/2$)~+~CaH($v=1,N=0, M_S=1/2$)  collisions, leading to efficient collisional cooling of CaH vibrations. 
\\
\end{abstract}

\maketitle

\section{Introduction}

The quantum dynamics of ultracold molecular collisions is a focal point of several major avenues of research within the emerging field of ultracold molecular gases \cite{Carr:09,Balakrishnan:16}.  In addition to their fundamental importance in    chemical physics, molecular collisions and chemical reactions determine many key properties of ultracold molecular gases, such as  thermalization rates, collision lifetimes, and controllability with external magnetic fields via magnetic Feshbach resonances.
A variety of intriguing quantum phenomena reveal themselves at ultralow temperatures due to the large de Broglie wavelengths of the colliding molecules \cite{Herschbach:09} and the absence of averaging over millions of quantum states \cite{Bohn:17}. Examples include threshold scattering, shape and orbiting resonances, tunnelling under the reaction barrier, quantum statistics, and  geometric phase \cite{Kendrick:15}, all of which can have profound,  largely unexplored, and potentially useful effects on collision rates at ultralow temperatures \cite{Balakrishnan:16}.

Recent experimental studies have explored how electric fields and quantum statistics affect the chemical reaction of two   KRb molecules at a temperature of 50 nK \cite{Ni:10,Ospelkaus:10,Miranda:11}, and of resonance scattering in cold He-NO \cite{Vogels:15} and  He$^*$-H$_2$ collisions \cite{Klein:16}.
In addition, rapid progress in molecular laser cooling \cite{Barry:14,McCarron:18,Truppe:17,Kozyryev:17,Anderegg:18} has opened the door to  studying ultracold collisions of molecules bearing unpaired electron spins  (such as CaH, SrF,  CaF, CaOH,  SrOH, and YbOH) with trapped ultracold atoms. 
Sympathetic cooling occurs through elastic (momentum-transfer) collisions when a gas of molecules to be cooled is put in thermal contact with an ultracold atomic ensemble \cite{Carr:09,Lara:06,Tscherbul:11,Morita:18}. 
Inelastic collisions decrease the efficiency of sympathetic cooling  by releasing the energy stored in internal molecular degrees of freedom. Therefore, sympathetic cooling can be expected to be effective when elastic collisions greatly outnumber inelastic collisions, with the ratio of elastic to inelastic cross sections $\gamma=\sigma_\text{el}/\sigma_\text{inel}>100$ \cite{Burke:99}. Very recently, sympathetic cooling has been observed experimentally in a  trapped mixture of NaLi$(^3\Sigma)$ molecules with Na($^2$S) atoms \cite{Son:20}.

These experimental advances strongly motivate rigorous, full-dimensional quantum dynamical calculations on ultracold molecular collisions in  the presence of external magnetic fields. When converged with respect to all basis set parameters, such calculations provide a direct connection between the scattering observables and the underlying potential energy surfaces (PES),  which could be used to map out intermolecular interactions  \cite{Campbell:09,Klein:16} and to provide direct information on the positions and widths of magnetic Feshbach resonances, which are responsible for quantum chaotic behavior in ultracold molecular collisions \cite{Croft:14,Frisch:14}.
Even in the presence of substantial uncertainties in the PESs, quantum scattering calculations can provide important statistical insights into  ultracold molecular scattering, such as the probability distributions of elastic and inelastic collision rates \cite{Morita:19b}.

However, rigorous quantum scattering calculations on ultracold atom-molecule collisions of current experimental interest (such as Li~+~CaH, Rb~+~SrF, and Rb~+~CaF) pose a challenging computational problem due to strong and anisotropic atom-molecule interactions, which  couple a large number of rotational states, leading to very large, computationally intractable systems of coupled-channel (CC) Schr\"odinger equations. The rotational basis set convergence problem can be partially overcome by using the computationally efficient total angular momentum representation for molecular collisions in external fields \cite{Tscherbul:10,Tscherbul:12}, which allows for converged computations with hundreds of rotational states \cite{Morita:17,Morita:18}. To save computational effort, these calculations kept the internuclear distance of the diatomic molecule fixed at its equilibrium value (the rigid rotor approximation), thereby neglecting the vibrational degrees of freedom (DOF).

Vibrational DOF are unique to molecular species, and can affect ultracold collision dynamics by causing vibrational relaxation (or quenching), inducing shape and/or Feshbach resonances \cite{Balakrishnan:98}, and modifying chemical reaction probabilities \cite{Balakrishnan:98,Liu:15}. 
Kozyryev {\it et al.} measured the cross section for vibrational relaxation of the first excited SrOH stretching mode  (1,0,0) in cold collisions with He atoms at 2~K and found it to be 700 times smaller than the diffusion cross section \cite{Kozyryev:15}. A large ratio of momentum transfer to vibrational relaxation cross sections ($\gamma = 10^4$) was measured for cold He~+~ThO($v=1,N=0$) collisions \cite{Au:14}.   Cold two-body collisions and three-body recombination of large molecules such as naphthalene, stilbene, and benzonitrile may be affected by their vibrational DOF \cite{Patterson:10,Piskorski:14,Putzke:12,Li:14}. Most recently, the prospects for laser cooling of complex polyatomic molecules with numerous vibrational modes have been explored~\cite{Li:2019,Klos:2020}. Understanding the role of vibrational DOF in ultracold collisions of such molecules is a prerequisite to efficient sympathetic cooling.

Previous theoretical work has explored the effects of  molecular vibrations on ultracold atom-molecule collisions in the absence of external fields using converged coupled-channel (CC) quantum scattering calculations. Balakrishnan {\it et al.} \cite{Balakrishnan:98}  studied ultracold He~+~H$_2$ collisions and observed rapid enhancement of vibrational relaxation with increasing the degree of vibrational excitation of H$_2$.  Balakrishnan and Dalgarno \cite{Balakrishnan:01b} and Volpi and Bohn \cite{Volpi:03} found that vibrational and fine-structure relaxation in ultracold He~+~O$_2$ collisions occur slowly, suggesting the possibility of buffer gas cooling and magnetic trapping of vibrationally excited O$_2$ molecules. Similar conclusions were reached by Balakrishnan, Krems and Groenenboom in their study of cold He~+~CaH($^2\Sigma$)  collisions \cite{Balakrishnan:03}.  More recent studies explored full-dimensional quantum dynamics of CO~+~H$_2$ collisions of astrophysical interest  \cite{Yang:15} and  found large vibrational cooling rates in cold Ca~+~BaCl$^+$ collisions \cite{Rellergert:13}.

Here, we report on the first rigorous, full-dimensional quantum scattering calculation on an atom-molecule collision in the presence of an external magnetic field, using Li~+~CaH as a representative example. The triplet Li-CaH interaction is strongly anisotropic \cite{Tscherbul:11} with a well depth of 0.7 eV, as is typical for polar $^2\Sigma$ molecules interacting with alkali-metal atoms \cite{Kosicki:17}. The enormous number of strongly coupled rovibrational basis states of the Li-CaH collision complex has thus far prevented numerically exact simulations of quantum dynamics of Li~+CaH collisions in external fields. In this work, we employ the computationally efficient total angular momentum basis \cite{Tscherbul:10} and make explicit use of inversion symmetry  to minimize the number of CC basis states. 
These improvements allow us to  achieve numerical convergence of Li~+~CaH scattering observables using extended rovibrational basis sets including up to 14
vibrational and 56 rotational states of CaH.

Armed with the improved methodology, we explore the collisional properties of an ultracold spin-polarized Li-CaH mixture in the presence of an external magnetic field. In particular, we test the rigid-rotor approximation employed in our previous calculations \cite{Morita:17,Morita:18} and find it to be qualitatively accurate, yet unable to reproduce the quantitative features of the inelastic cross sections.  We also  study vibrational energy transfer from the $v=1$ vibrational level of CaH induced by ultracold collisions with Li atoms. Our calculations reveal rapid, magnetic field-insensitive vibrational quenching in ultracold Li~+~CaH($v=0,N=0$) collisions,  implying efficient  collisional cooling of the vibrational DOF of CaH.

This paper is organized as follows. In Sec. II we present our  three-dimensional {\it ab initio} PES of the Li-CaH collision complex. In Sec. IIB we outline the details of quantum scattering calculations of  ultracold Li~+~ CaH collisions, focusing on the new aspects of including the  dependence on the CaH vibrational coordinate. The results of quantum scattering calculations are presented and discussed in Sec.~III, whereas Sec. IV summarizes the main results  of this work.

 \color{Black}

\section{Theory}

\begin{figure*}[t]
\begin{center}
\includegraphics[width=\textwidth]{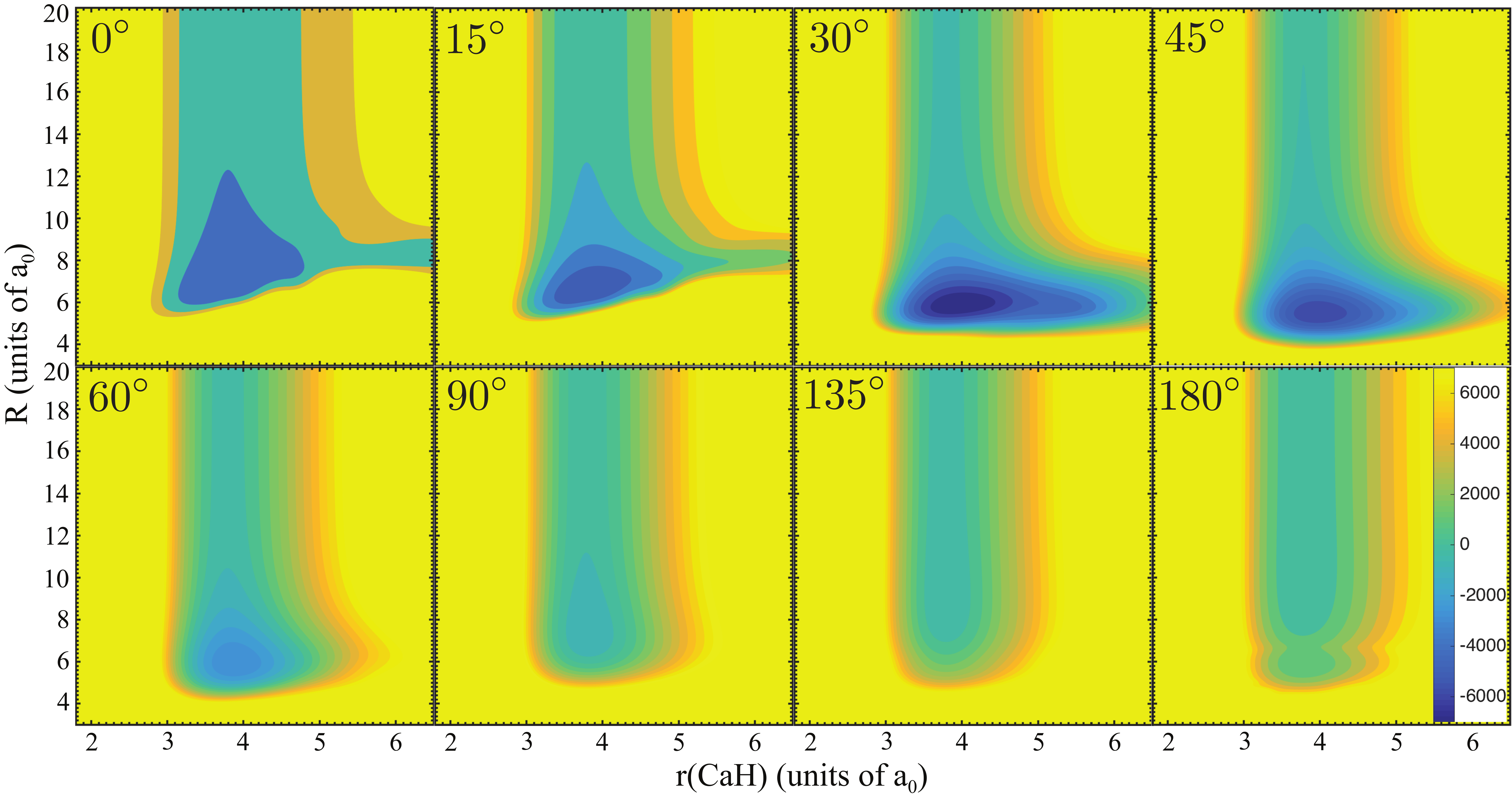}
\end{center}
\caption{Contour plot representing the 3D Li-CaH PES of triplet symmetry as a function of the Jacobi coordinates $r$, $R$, and $\theta$. The global minimum of -7202.4 cm$^{-1}$ is located at $r_e=3.960$ a$_0$, $R_e=5.804$ a$_0$ and $\theta_e=32.5^o$. The values of the PES are given in the color map (bottom right) in units of cm$^{-1}$. Lines of the same color represent contours of equal energy.  }
\label{fig_3DPES}
\end{figure*}

\subsection{{Ab initio} interaction PES}

As a prerequisite to full-dimensional quantum dynamical calculations of ultracold Li~+~CaH collisions, we have computed a new three-dimensional (3D) {\it ab initio} interaction PES for the triplet electronic state of the Li-CaH collision complex. To this end, we carried out state-of-the-art electronic structure calculations of the Li-CaH interaction energy in Jacobi coordinates ($R$, $r$, $\theta$), where $R$ is the distance between Li and the center of mass of CaH, $r$ is the bond distance of CaH and $\theta$ is the angle between the Jacobi vectors $\bf{R}$ and $\bf{r}$. All of the {\it ab initio} calculations of the two-body CaH and three body Li-CaH potentials were performed with the unrestricted coupled-cluster method with single, double and noniterative triple excitations, UCCSD(T). To include the effects of the CaH vibrational motion, its internuclear distance was varied from $r= 2.27\,a_0$ to $r=6.62\,a_0$, which is sufficiently wide to cover the range of CaH vibrational states $v=0{-}13$.  The value $r=6.62\, a_0$ is the largest one accessible with the single-reference UCCSD(T) method, as further stretching resulted in an unphysical interaction potential and  convergence problems due to the increasingly multireference character of the electronic wavefunction. We used the same UCCSD(T)  method and basis set of quadruple-zeta quality augmented with mid-bond functions as in our previous two-dimensional rigid-rotor calculations \cite{Tscherbul:11}. 

Figure~\ref{fig_3DPES} shows a series of contour plots of our Li-CaH PES as a function of $R$ and $r$ at fixed  values of $\theta$. The PESs are obtained by adding the 3D interaction potential $V(r,R,\theta)$  to the two-body CaH potential $V_\text{CaH}(r)$.   The global minimum of the PES is 7202.4 cm$^{-1}$ deep and is located at a skewed Li-CaH geometry ($r_e=3.960\,a_0$, $R_e=5.804\,a_0$, and $\theta_e=32.5^\text{o}$). The PES becomes increasingly more  repulsive with increasing $\theta$ as shown in Fig.~\ref{fig_3DPES}. We also note that the triplet PES, unlike the singlet one \cite{Tscherbul:20}, is not reactive within the range of geometries considered.

Our 3D PES is slightly  more attractive than the 2D rigid-rotor PES developed in our previous calculations \cite{Tscherbul:11}, which is characterized by $r_e=3.803\,a_0$, $R_e=5.6\, a_0$, $\theta=35.0^o$ and by the well depth of 7063 cm$^{-1}$. The inclusion of the CaH stretch pushes the global minimum to larger $R$ by $\simeq$0.2 $a_0$, increases the equilibrium bending angle by 3 degrees, and deepens the well by  $\simeq$160 cm$^{-1}$. 

In this work, we are interested in  collisions of Li and CaH in their maximally spin-stretched initial states $M_{S_A}=M_{S_B}=1/2$ and $M_{S_A}=M_{S_B}=-1/2$. Neglecting weak spin-dependent interactions \cite{Abrahamsson:07,Tscherbul:20}, the total spin of the collision complex  $S$ remains a good quantum number, so we only need to consider the Li-CaH PES of the triplet symmetry ($S=1$).

\subsection{Quantum scattering theory and computational methodology}

Our full-dimensional quantum scattering calculations are based on the exact solution of the time-independent Schr\"odinger equation for a $^2\Sigma^+$ molecule ($A$) colliding with a $^2$S atom ($B$) in an external magnetic field. The details of the theoretical formalism and its numerical implementation were given in our previous work   \cite{Morita:18,Tscherbul:11,Morita:19b, Morita:17}, so we will omit the material already covered there  and focus on the aspects of the quantum scattering methodology pertinent to  vibrational DOF.

The atom-molecule collision complex is described by the Hamiltonian 
\begin{equation}
\hat{\mathcal{H}} = - \frac{1}{2\mu R}  \frac{d^2}{dR^2} R
                     + \frac{\hat{L}^2}{2\mu R^2} \allowbreak
                     + \hat{\mathcal{H}}_{{A}}
                     + \hat{\mathcal{H}}_{B}
                     + \hat{\mathcal{H}}_{\mathrm{int}},
\label{eq:Heff}
\end{equation}
where  $R$ is the atom-molecule distance, $\hat{L} = \hat{J}-\hat{N}-\hat{S}_{{A}}-\hat{S}_{{B}}$ is the orbital angular momentum of the collision complex with reduced mass $\mu$ expressed through its  total angular momentum $\hat{J}$, the molecular rotational angular momentum $\hat{N}$, and the atomic and molecular electron spin angular momenta $\hat{S}_i$ ($i={A},B$).  
The Hamiltonian of the vibrating $^2\Sigma^+$ diatomic molecule is  
\begin{equation}\label{Hmol}
\hat{\mathcal{H}}_{A} = - \frac{1}{2\mu_A r}  \frac{d^2}{dr^2}r + \frac{\hat{N}^2}{2\mu_A r^2}  + V_A(r) + \gamma_{sr} \hat{N} \cdot \hat{S}_{A} + \hat{H}_{AZ}
\end{equation}
where $\mu_{A}$ is the reduced mass of the molecule,  ${r}$ is the internuclear distance, $V_A(r)$ is the potential energy curve for the $X ^2\Sigma^+$ molecular ground state, and $\gamma_{sr}$ is the spin-rotation constant. 
The interaction of collision partners with the external magnetic field is described by the corresponding Zeeman Hamiltonians  $\hat{H}_{AZ} = g_e\mu_B \hat{S}_{A_Z} B$ and $\hat{\mathcal{H}}_{BZ}  = g_e\mu_B \hat{S}_{{B}_Z} B$,
  where $\hat{S}_{i_Z}$ are the projections of $\hat{S}_{i}$ onto the magnetic field axis, $g_e\simeq 2.002$ is the electron $g$-factor, and $\mu_{B}$ is the Bohr magneton.  The atomic Hamiltonian $\hat{\mathcal{H}}_{B}=\hat{\mathcal{H}}_{BZ}$ since we neglect the atomic and molecular hyperfine structure known to play a minor role in spin-polarized collisions considered here \cite{Morita:18}.
The atom-molecule interaction is given by
\begin{equation}\label{Hint}
\hat{\mathcal{H}}_{\mathrm{int}}= \sum_{S,M_S} {V}^S(R,r,\theta)|SM_S\rangle\langle S M_S| +\hat{{V}}_{dd}(\bf{R})
\end{equation}
 where ${V}^S(R,r,\theta)$ are the 3D interaction PESs for the singlet $(S=0)$ and triplet ($S=1$) electronic states of the Li-CaH complex.  Here, we consider ultracold collisions between  Li and CaH in their fully spin-polarized initial states and we neglect the small matrix elements of the magnetic dipole-dipole interaction $\hat{{V}}_{dd}(\bf{R})$ \cite{Morita:17,Janssen:11, Morita:19b} and of the  spin-rotation interaction $\gamma_{sr} \hat{N} \cdot \hat{S}_{A}$ (\ref{Hmol}) between the states of different $S$. In this approximation,  spin-polarized collisions occur exclusively on a single $S=1$ PES \cite{Abrahamsson:07}  calculated  {\it ab initio}  as described in Sec. IIA.

To solve the time-independent Schr\"{o}dinger equation, we expand the total wavefunction of the collision complex in a set of spin-rovibrational basis functions \cite{Tscherbul:10}  
\begin{equation}
\label{eq:totjbasis}
|\Psi\rangle = \frac{1}{R} \sum^{}_{\alpha,J,\Omega}
F^{Mp}_{\alpha J \Omega}(R) |\Phi_{\alpha J \Omega}^{Mp}\rangle 
\end{equation}
where $F^{Mp}_{\alpha J \Omega}(R)$ are the radial expansion coefficients and  the parity-adapted body-fixed (BF) channel basis states are given by  
\begin{equation}
\begin{split}
& |\Phi_{\alpha J \Omega}^{Mp}\rangle =\frac{1}{\sqrt{2}} |\chi^N_v\rangle [|JM\Omega\rangle  |NK_N\rangle|S_{{A}}\Sigma_{{A}}\rangle|S_{{B}}\Sigma_{{B}}\rangle \\ +\ & (-1)^{X+p}|JM-\Omega\rangle |N-K_N\rangle|S_{{A}}-\Sigma_{{A}}\rangle|S_{{B}}-\Sigma_{{B}}\rangle],
\label{eq:parity}
\end{split}
\end{equation}
where $p=\pm1$ gives the inversion parity, $X=J-S_{A}-S_{B}$,  and $|\chi^N_v\rangle$ are the radial part of the rovibrational eigenfunctions of the diatomic molecule defined below. The angular basis functions  are composed of the eigenstates of $\hat{N}^2$, $\hat{N}_z$, $\hat{S}_i^2$, and $\hat{S}_{i_z}$ ($i=A,B$), where 
$K_N$, $\Sigma_{{A}}$ and $\Sigma_{{B}}$ (inlcuded in the collective index  $\alpha$) are the projections of $N$, $S_{{A}}$, $S_{{B}}$ onto the BF quantization $z$ axis defined by the atom-molecule Jacobi vector $\bf{R}$ \cite{Tscherbul:10,Morita:17,Janssen:11, Morita:18}.
Further, $|JM\Omega\rangle$ are symmetric top eigenfunctions \cite{Zare:88}  parametrized by the Euler angles, which specify the orientation of the BF coordinate axes in the space-fixed (SF) frame \cite{Tscherbul:10,Morita:17}, $\Omega$ is the projection of $J$ onto the BF $z$ axis, $\Omega=K_N+\Sigma_{A}+\Sigma_{B}$.
We note that $M$, the projection of $J$ onto the SF quantization axis, is rigorously conserved for collisions in a static magnetic field.  Here, we are interested in collisions of spin-1/2 particles, so both  $\Sigma_{A}$ and $\Sigma_{B}$ are nonzero and the normalization factor required in Eq. (\ref{eq:parity}) for $K_N=\Sigma_{A}=\Sigma_{B}=0$ can be omitted.

The asymptotic behavior of the radial coefficients $F^{M p}_{\alpha J\Omega}(R)$ defines the scattering $S$-matrix, which  provides a complete microscopic description of quantum scattering \cite{Nock:14}. 
To find the coefficients $F^{Mp}_{\alpha J\Omega}(R)$, we substitute the CC expansion (\ref{eq:totjbasis}) to the time-independent Schr{\"o}dinger equation $\hat{H}\Psi=E\Psi$, where $E$ is the total energy, to obtain the CC equations \cite{Secrest:79,Tscherbul:10,Tscherbul:18b}. 
\begin{multline}\label{eq:CC}
\Bigl{ [ }  \frac{1}{2\mu }\frac{d^2}{dR^2}\ + E \Bigr{ ] } F^{Mp}_{\alpha J\Omega}(R) \\
= \sum^{}_{\alpha',J',\Omega'} 
\langle \Phi_{\alpha J \Omega}^{Mp} | \frac{\hat{L}^2}{2\mu R^2}  
+ \hat{\mathcal{H}}_{{A}}+ \hat{\mathcal{H}}_{B}+ \hat{\mathcal{H}}_{\mathrm{int}}
| \Phi_{\alpha' J'\Omega'}^{Mp}\rangle F^{Mp}_{\alpha' J'\Omega'}(R) .
\end{multline}
The matrix elements over the angular basis functions can be evaluated analytically as described elsewhere  \cite{Tscherbul:10,Pack:74, Morita:17, Morita:18}.
As a first step, we expand the 3D Li-CaH interaction PES in Legendre polynomials 
 \begin{equation}
 V^S(R,r,\theta)=  \sum^{\lambda_\text{max}}_{\lambda=0}  V^{S}_\lambda(R,r) P_{\lambda}(\cos\theta).
\label{eq:V_expand}
\end{equation}
with $\lambda_\text{max}=14$. 
The explicit inclusion of the vibrational DOF of CaH  $r$  is an essential new element of this work,  which distinguishes it from previous  atom-molecule scattering calculations in a magnetic field based on the rigid-rotor approximation \cite{Tscherbul:18b}.

The matrix elements over the rovibrational basis functions involve the vibrational integrals \cite{Tscherbul:10, Morita:18,Krems:04} 
 \begin{equation}
 \langle \chi^{N'}_{v'}| V^{S}_\lambda(R,r) |\chi^N_v\rangle =    \int_0^\infty {\chi^{N'}_{v'}}(r) V^{S}_\lambda(R,r) \chi^N_v(r)\ dr,
\label{eq:V_lambda}
\end{equation}
where the vibrational basis functions  $\langle r|\chi^N_v\rangle=r^{-1}\chi^N_v(r)$
in \cref{eq:parity} are solutions of the one-dimensional  Schr{\"o}dinger equation 
\begin{equation}
\Bigl{ [ } \frac{-1}{2\mu_{A}} \frac{d^2}{dr^2} + \frac{N(N+1)}{2\mu_{A} r^2} +V_{A}(r) \Bigr {] } \chi^N_v(r) = E_{vN} \chi^N_v(r).
\label{eq:mol}
\end{equation}
where $E_{vN}$ are the rovibrational energy levels of the isolated CaH molecule.

We solve \cref{eq:mol} numerically for each value of $N$   by diagonalizing the Hamiltonian in the basis of particle-in-a-box discrete variable representation (sinc-DVR) basis functions \cite{Colbert:92}. A total of 120 DVR basis functions are employed to obtain converged bound rovibrational levels of CaH with $v\leq 13$.
We use the value $\mu_{A}=0.9830336974$ amu for the reduced mass of CaH. 
The vibrational integrals in \cref{eq:V_lambda} are pre-evaluated on a Gauss-Legendre quadrature and stored on hard disk for subsequent use in scattering calculations.

The size of the CC basis set in \cref{eq:totjbasis} is controlled by the cutoff parameters $J_\mathrm{max}$, $v_\mathrm{max}$, and $N_\mathrm{max}$, which give the maximum values of $J$, $v$, and $N$. We use $J_\text{max}=1$, $N_\text{max}=55$ in all calculations, and fix the total angular momentum projection to  $M=M_{S_A}+M_{S_B}$, which corresponds to the incident $s$-wave channel.  The convergence of scattering results with respect to $v_\text{max}$ is examined in the next section.  The total number of scattering channels $N= 4634$ for $v_\text{max}=13$ and $N = 331$ for  $v_\text{max}=0$.
Most of the parameters and physical constants employed in the CC calculations are the same as those used in our previous work \cite{Tscherbul:10,Morita:19b}.

The  CC equations are solved numerically by means of the Johnson-Manolopoulos  log-derivative algorithm  \cite{Manolopoulos:86}. To save computational effort, we separated the radial grid into two regions 
 extending from  $R_\text{min}=4.0\, a_0$ to $R_\text{mid}=20.0\, a_0$ (inner region) and from $R_\text{mid}$ to $R_\text{max}$ (outer region), with the grid spacing of $0.01\,a_0$ in the inner region and 0.1~$a_0$ in the outer region. The maximum value of $R_\text{max}$ used in the present calculations is $1500\,a_0$.
 At $R=R_\text{max}$, the log-derivative matrix is transformed from the total angular momentum representation   \cref{eq:totjbasis} to a basis set which diagonalizes $\hat{H}_A$, $ \hat{\mathcal{H}}_{BZ}$, and the square of the orbital angular momentum operator $\hat{L}^2\  =(\hat{J}-\hat{N}-\hat{S}_{{A}}-\hat{S}_{{B}})^2$   \cite{Tscherbul:10,Morita:18}. The transformed log-derivative matrix is then matched to the scattering boundary conditions to obtain the $S$-matrix and scattering cross sections \cite{Tscherbul:10,Morita:18}.

\color{black}


\section{Results}


\subsection{Vibrational basis set convergence}

\begin{figure}[t!]
\begin{center}
\includegraphics[width=\linewidth]{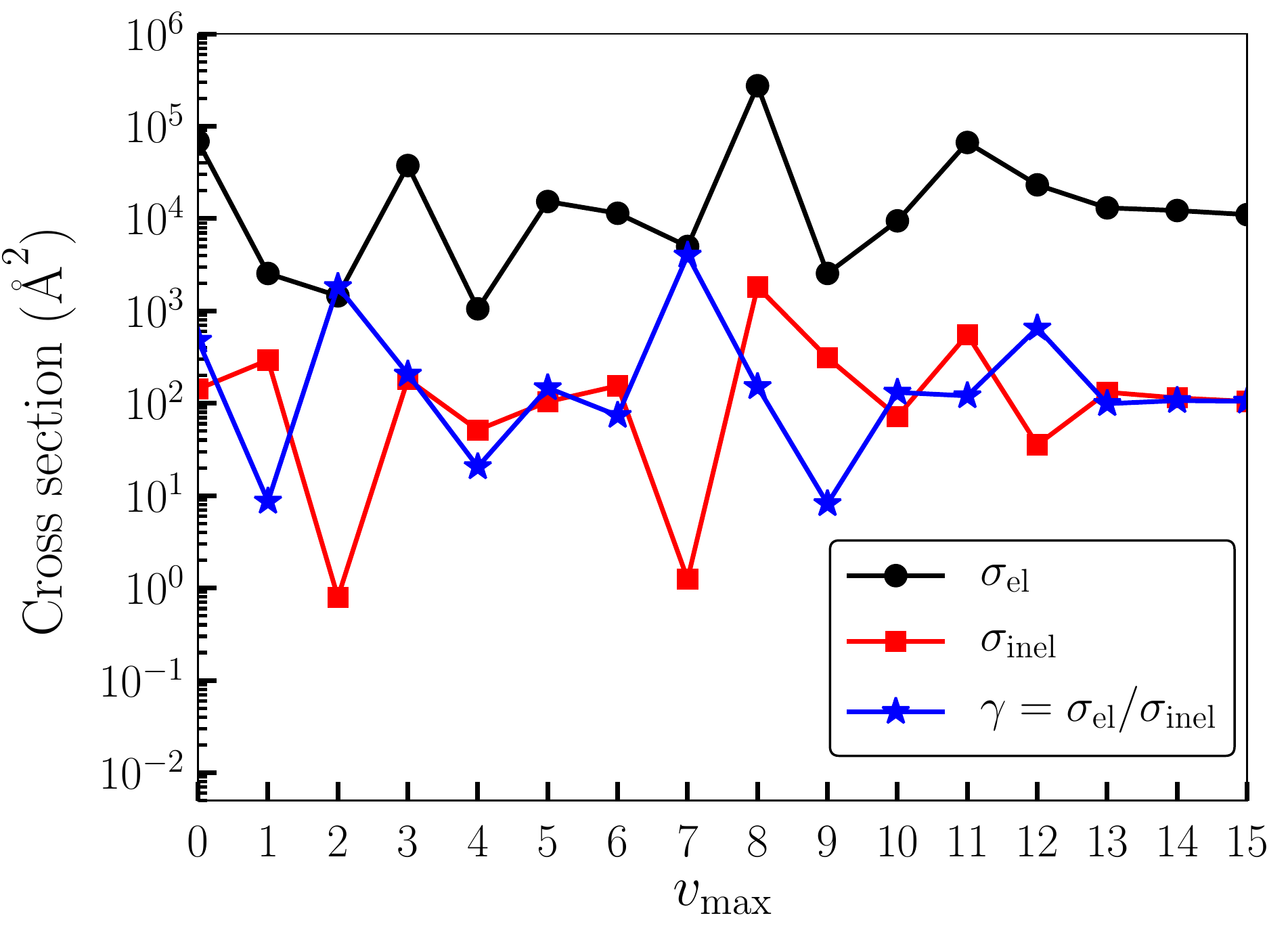}
\end{center}
\caption{Convergence of elastic cross section (black circles), inelastic cross section (red squares) and elastic-to-inelastic ratio (blue stars) for spin-polarized Li+CaH ($v=0$, $N=0$) collisions with respect to the highest vibrational state of CaH included in the basis set at the collision
energy of $E_C=10^{-6}$ cm$^{-1}$. The magnitude of magnetic field is 1000 G, $J_\text{max} = 1$, and  $N_\text{max}=55$.
}
\label{fig_Conv_vmax}
\end{figure}

To obtain numerically exact collision observables, it is essential to examine their dependence on CC basis set size and to demonstrate numerical convergence. Only the results that are converged (or nearly converged) with respect to basis set correspond to physically meaningful solutions of the Schr\"odinger equation for a given PES \cite{Secrest:79,Suleimanov:16,Tscherbul:18b}. We note that basis set convergence is not required when one in interested in quantities averaged over an ensemble of PESs, 
 such as the cumulative probability distributions \cite{Morita:19b} examined in Sec. IIID below.

\Cref{fig_Conv_vmax} shows the elastic and inelastic cross sections as a function of $v_\text{max}$, the maximum vibrational state included in the CC basis (with each vibrational manifold containing 56 rotational states to ensure rotational convergence, see Appendix A). We observe the onset of convergence at  $v_\text{max}=10$, with both the elastic and inelastic cross sections reaching their limiting values at $v_\text{max}\ge 13$.   The data shown in \cref{fig_Conv_vmax}  represent the first numerically converged quantum scattering results for a  strongly anisotropic atom-molecule collision system in three dimensions in the presence of an external magnetic field.

 Significantly, as shown in \cref{fig_Conv_vmax}, it is necessary to include as many as 14 vibrational states ($v = 0{-}13$) to fully saturate the basis set.  Such a slow convergence is due to the strong vibrational anisotropy of the Li-CaH PES  illustrated in \cref{fig_3DPES}.  Indeed, the bond distance of CaH near the global minimum structure of the Li-CaH complex is significantly enlarged with respect to the equilibrium bond length, leading to strong couplings between the different vibrational states.
 A similarly slow convergence is observed with respect to rotational basis set size \cite{Tscherbul:11,Suleimanov:16} due to   the strong coupling between the molecular rotational states induced by the angular  anisotropy of the PES. 
All our subsequent calculations will employ the fully converged $v_\text{max}=13,\,N_\text{max}=55$ basis unless stated otherwise.

\subsection{Ultracold collision dynamics of Li~+~CaH($v=0$) in the ground vibrational state}

\begin{figure}[t!]
\begin{center}
\includegraphics[width=\linewidth]{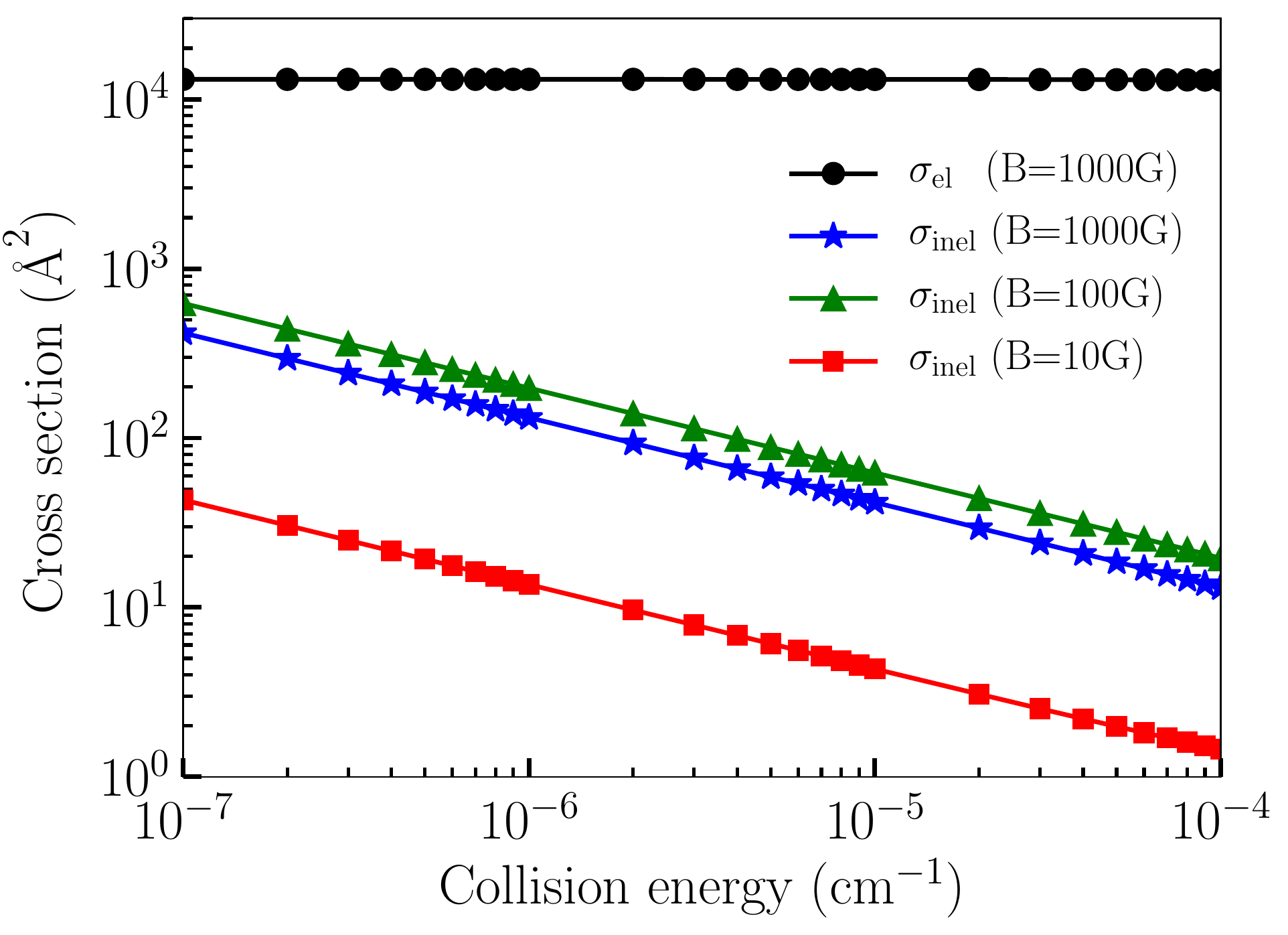}
\end{center}
\caption{Collision energy dependence of elastic cross section (black circles) and inelastic spin-relaxation cross sections for the external magnetic field of 1000 G (blue stars), 100 G (green crosses), and 10 G (red squares) in spin-polarized Li-CaH ($v=0$, $N=0$) collisions. The elastic cross section displays a very weak magnetic field dependence.
}
\label{fig_CSvsE}
\end{figure}

We will begin by focusing on ultracold collisions of CaH molecules in their ground electronic and rovibrational state  ($v=0, N=0$)  with Li atoms in their ground electronic state, with both Li and CaH in their maximally spin-stretched Zeeman states ($M_{S_{A}} = M_{S_{B}} = 1/2$). Here, $M_{S_{A}}$ and $M_{S_{B}}$ are the projections of $\hat{S}_{A}$ and $\hat{S}_{B}$ onto the space-fixed quantization axis defined by the external magnetic field.
Below, we will  be interested in elastic (inelastic) collisions, which conserve (change)   the values of $M_{S_{A}}$  and $ M_{S_{B}}$. As only low-field-seeking states with $M_{S_i}=+1/2$ can be confined in a magnetic trap, inelastic  collisions lead to undesirable trap loss \cite{Carr:09,Morita:18}

\Cref{fig_CSvsE} shows the cross sections for elastic scattering and inelastic spin relaxation in Li($M_{S}=1/2$)~+~CaH$(v=0,N=0,M_{S}=1/2)$ collisions as a function of collision energy. Following the expected  $s$-wave Wigner threshold  behavior, the elastic cross sections approach a constant value and the inelastic cross sections scale as $E_C^{-1/2}$, where $E_C$ is the collision energy.  We observe that elastic collisions occur much faster than inelastic collisions, indicating good prospects for sympathetic cooling of spin-polarized CaH molecules   with Li  atoms in a magnetic trap. The predominance of elastic over inelastic scattering was also observed in our previous reduced-dimensional  calculations  on spin-polarized Li~+~CaH collisions \cite{Tscherbul:11}, suggesting that the rigid-rotor approximation can provide a qualitatively correct picture of spin relaxation and elastic scattering in ultracold collisions of $^2\Sigma$ molecules in their ground vibrational states. However, as shown below, the accurate treatment of vibrational DOF is essential for the correct description of the magnetic field dependence of scattering cross sections, as well as for describing ultracold collisions of vibrationally excited CaH molecules with Li atoms.

\begin{figure}[t!]
\begin{center}
\includegraphics[width=\linewidth]{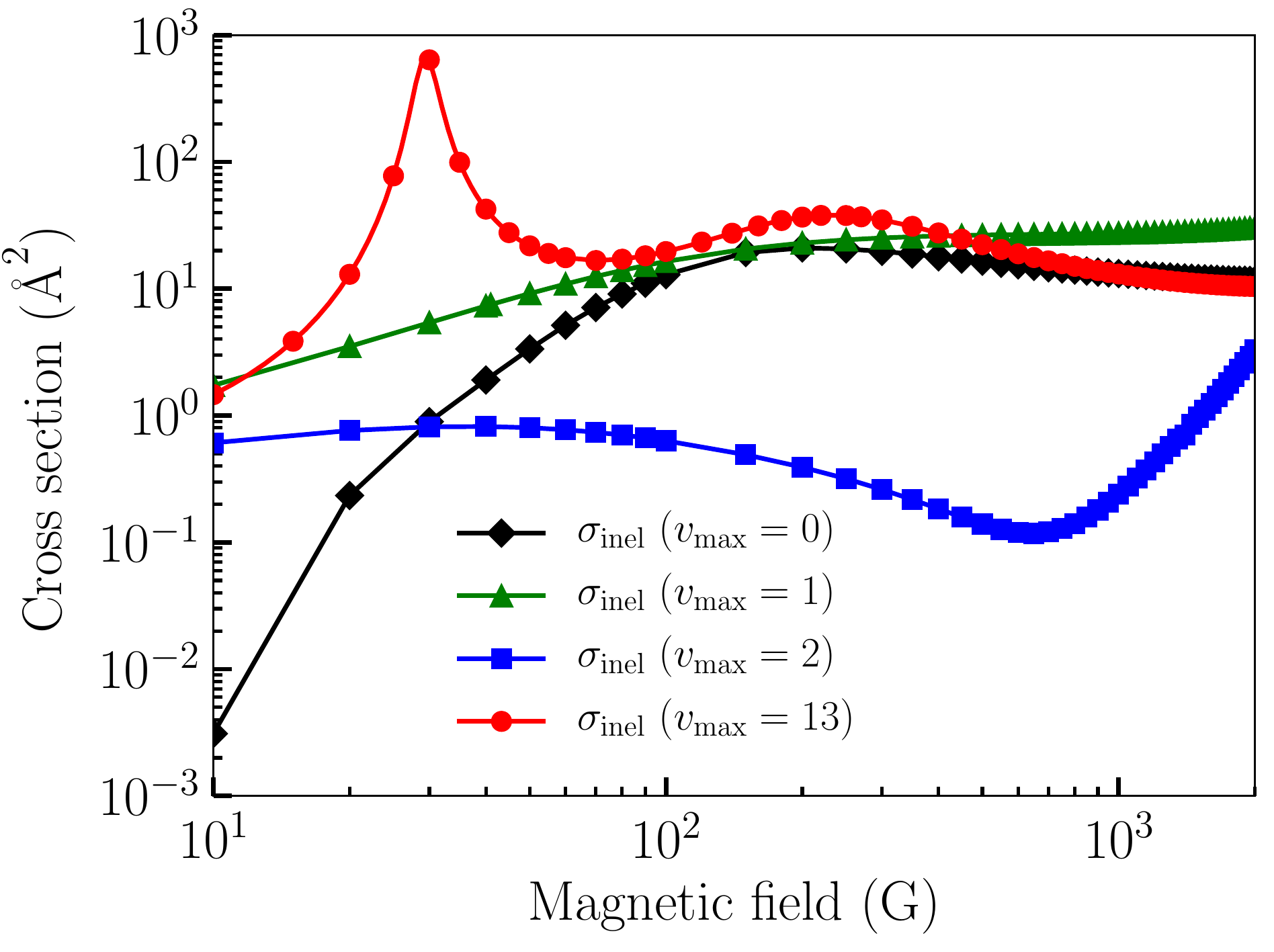}
\end{center}
\caption{
Magnetic-field dependence of inelastic spin-relaxation cross section in spin-polarized Li-CaH ($v=0$, $N=0$) collisions at the collision energy of $10^{-4}$ cm$^{-1}$ with $v_\mathrm{max}=0$ (black diamonds), $v_\mathrm{max}=1$ (green triangle), $v_\mathrm{max}=2$ (blue squares) and $v_\mathrm{max}=13$ (red crosses).
}
\label{fig_CSvsB}
\end{figure}

In \cref{fig_CSvsB} we plot the magnetic field dependence of the inelastic cross section at a collision energy of $10^{-4}$~cm$^{-1}=0.14$~mK. A pronounced resonance peak is observed  at $B\simeq 30$~G followed by a broad peak at $B\simeq 300$~G.
To explore whether the resonance peak could be reproduced in calculations using restricted basis sets,  we calculated the magnetic field dependence of the inelastic cross section using basis sets containing only the ground ($v_\text{max}=0$), the ground and the first excited ($v_\text{max}=1$), and the lowest three ($v_\text{max}=2$) vibrational states of CaH.
As shown in \cref{fig_CSvsB}, no low-field resonances  are observed in truncated basis calculations. Furthermore, as shown in Appendix B, the 30~G resonance cannot be reproduced by scaling the interaction PES by a constant factor $\lambda$. This suggests that  fully converged  three-dimensional calculations are required to map out the magnetic field dependence of the inelastic cross section. At higher magnetic fields, truncated basis calculations are in better agreement with the fully converged results. In particular, the broad resonance peak at $B\simeq 300$~G is well reproduced with only a single vibrational state in the basis ($v_\text{max}=0$).

We next explore the quantum dynamics of Li~+~CaH  collisions with both collision partners in their absolute ground states  ($M_{S_{A}} = M_{S_{B}} = -1/2$). Specifically, we are interested in how the vibrational DOF affect the density of magnetic Feshbach resonances, which could be used to control the scattering length in ultracold atom-molecule mixtures. The statistics of the resonance states determines the degree to which quantum chaos manifests itself in ultracold atom-molecule collisions  \cite{Frisch:14, Morita:16}, and the probability of complex formation (``sticking'') leading to three-body recombination and loss of trapped molecules \cite{Piskorski:14,Croft:14}.

\begin{figure}[t!]
\begin{center}
\includegraphics[width=\linewidth]{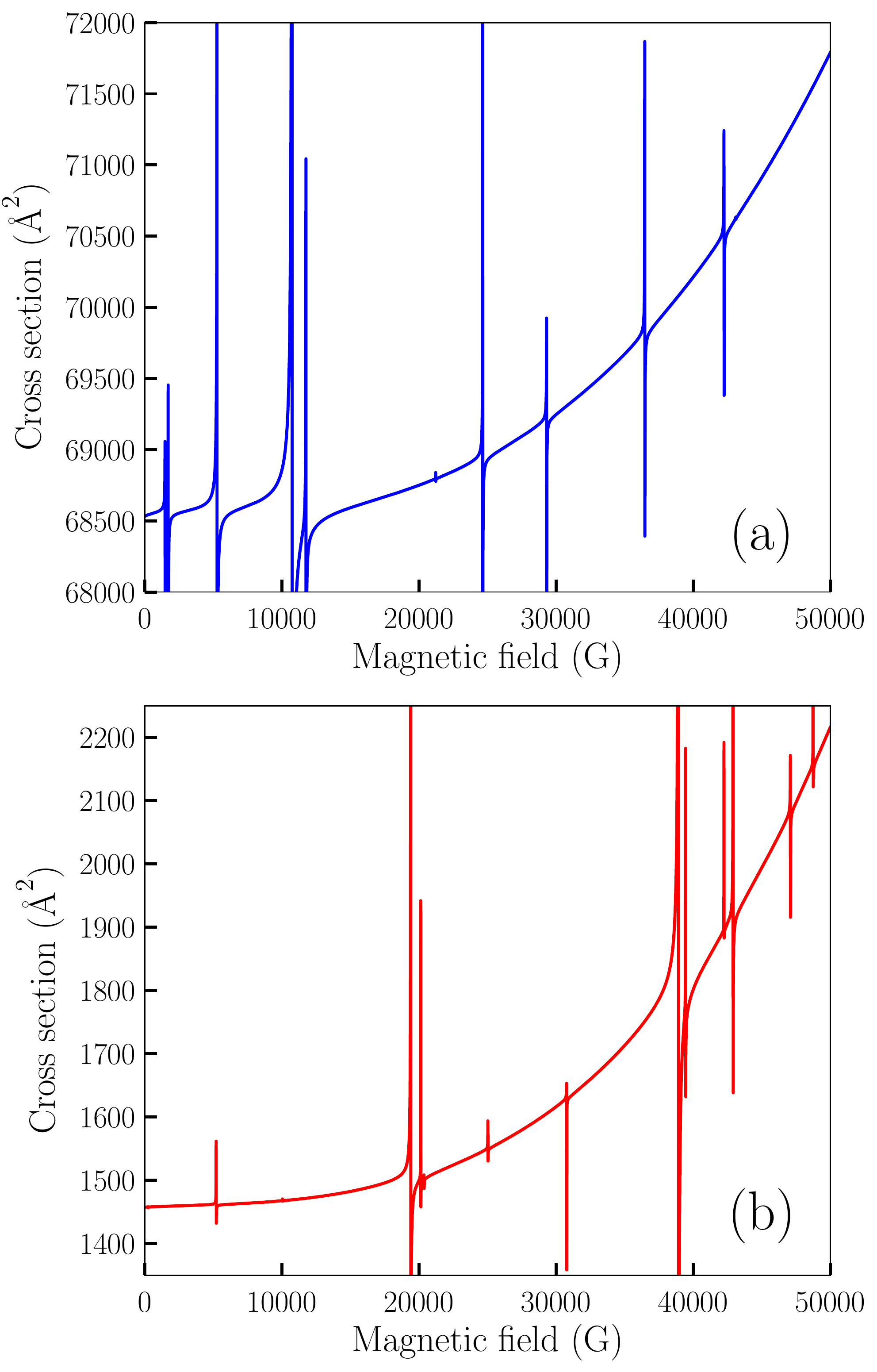}
\end{center}
\caption{Magnetic-field dependence of the elastic cross section for Li~+~CaH ($v=0$, $N=0$) collisions in the absolute ground Zeeman states at the collision energy of $10^{-6}$~cm$^{-1}$ (1.4 $\mu$K) calculated with $v_\mathrm{max}=0$ (a) and  $v_\mathrm{max}=2$ (b).
 }
\label{fig_CSvsB_gs}
\end{figure}

\Cref{fig_CSvsB_gs} shows the magnetic field dependence of the Li~+~CaH elastic cross section calculated with and without the vibrational DOF of CaH.  To locate narrow resonances, we used a dense magnetic field grid  of 5001 points  
spanning the range from 0 to 5$\times 10^4$ G ($\Delta B=10$ G). To make the calculations computationally feasible,  we truncated  the vibrational basis set to include three lowest-energy vibrational levels of CaH ($v_\text{max}=2$). By carrying out fully converged calculations over a limited range of fields ($B=0-0.2$~T) we verified that the inclusion of highly excited vibrational states with $v\ge 3$ does not qualitatively affect the number of resonances between 0 and 5 T.

\begin{figure}[t!]
\begin{center}
\includegraphics[width=1.05\linewidth, trim = 40 0 0 35]{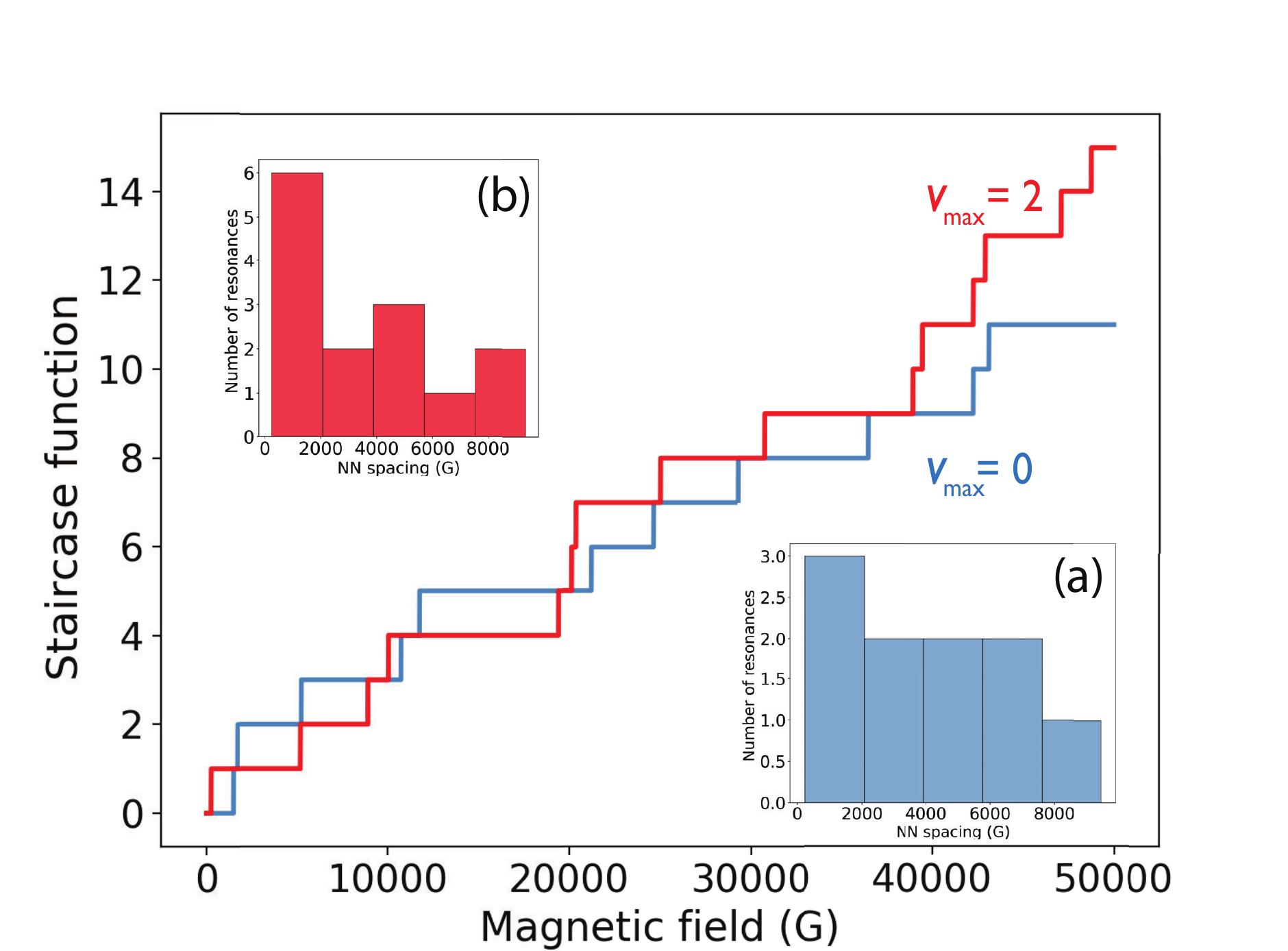}
\end{center}
\caption{
Magnetic field dependence of the staircase function $N(B)$ for the resonance spectra shown in \cref{fig_CSvsB_gs} calculated with $v_\mathrm{max}=0$ and  $v_\mathrm{max}=2$. The corresponding  nearest-neighbor spacing distributions are shown in  insets (a) [for $v_\text{max}=0$] and (b) [for $v_\text{max}=2$].}
\label{fig_staircase}
\end{figure} 

Comparing \cref{fig_CSvsB_gs} (a) and (b), we observe that the number of magnetic Feshbach resonances in Li~+~CaH collisions is fairly low, and it does not change significantly with increasing the size of the vibrational basis set. 
Within the broad range of $B = 0{-}5$~T, we observe a total of 11(15) resonances with $\Gamma > 10$~G for $v_\text{max}=0(2)$.

To estimate the mean density of resonances per 1~G ($\bar{\rho}$), we calculated the staircase function $N(B)$  defined as the total number of resonances below the magnetic field $B$ \cite{Frisch:14}. \Cref{fig_staircase} shows that  the staircase functions calculated for $v_\text{max}=0$ to $v_\text{max}=2$ are qualitatively similar at $B<4$~T. At higher fields the $v_\text{max}=2$ staircase function grows faster due to the larger number of resonances in this region for $ v_\text{max}=2$ as compared to  $v_\text{max}=0$ (see \cref{fig_CSvsB_gs}). The step-like behavior of the staircase functions indicates that the number of resonances identified in our calculations is insufficient to generate a large enough sample. Thus, the results presented below should be regarded as qualitative estimates, rather than statistically significant predictions.

 From a linear fit to $N(B)$ we determine the mean resonance density to be $2.0\times 10^{-4}$ G$^{-1}$ for $v_\text{max}=0$ and  $2.7\times 10^{-4}$~G$^{-1}$ for  $v_\text{max}=2$. The resonance density thus increases slightly with the number of vibrational states included in the basis set, although this increase is not nearly as significant as might be expected based on the number of states of the Li-CaH collision complex included in the basis (which grows by a factor of 3 in going from $v_\text{max}=0$ to $v_\text{max}=2$).  

The nearest-neighbor spacing (NNS) distributions of the resonance positions for $v_\text{max}=0$ and $v_\text{max}=2$ are plotted in the insets of \cref{fig_staircase}. As with the staircase functions, we observe that the distributions are qualitatively similar, indicating that including more vibrational levels in the basis set does not strongly affect the  NNS  distributions.

 \color{black}

Our results suggest that the mean resonance density is not strongly affected by the vibrational DOF of CaH. It should be noted that   intramolecular hyperfine interactions \cite{Tscherbul:07,GonzalezMartinez:11,Tscherbul:20}, omitted here, are nearly diagonal in vibrational quantum numbers (al least for low $v$). Therefore, while these interactions are likely to increase the overall  resonance density, they are unlikely to affect the way, in which it is influenced by vibrational DOF.

 To understand the physical origin of the low resonance density, we note that the atom-molecule scattering length in the vicinity of a magnetic Feshbach resonance may be written as \cite{Chin:10}
\begin{equation}\label{FR}
a(B)=a_\text{bg} - a_\text{bg}\frac{\Delta}{B-B_0},
\end{equation}
where $\Delta=\Gamma_0/\delta\mu$ is the zero-energy resonance width expressed via the coupling matrix element between the open and closed channels 
\begin{equation}\label{Gamma0}
\Gamma_0=2\pi |\langle \psi_c^{M_c}|\hat{W}|\psi_o^{M_o}(E_0)\rangle|^2,
\end{equation}
and $\Delta M=M_o -M_c$ is the difference between the magnetic quantum numbers of the open-channel state $|\psi_o^{M_o}(E_0)\rangle$ located above the atom-molecule threshold and the atom-molecule bound state $|\psi_c^{M_c}\rangle$. Note that $\Delta M \ne 0$ is required for these states to cross, leading to the appearance of a zero-energy magnetic Feshbach resonance. To couple the states of different $M$, the interaction $\hat{W}$ responsible for the resonance width in \cref{Gamma0} must be  spin-dependent.

Here, we are interested in ultracold collisions of $^2\Sigma$ molecules $^2$S atoms initially in their maximally spin-stretched Zeeman states (either high or low-field seeking). In this particular case, the atom-molecule interaction PES is spin-independent and does not couple the states of different $M$. As a result, magnetic Feshbach resonances can only originate from the intramolecular spin-rotation and intermolecular magnetic dipole-dipole interactions [\cref{Hmol,Hint})]. These interactions are diagonal in $v$, so {only $v'=0$ closed channels are expected to contribute to the magnetic couplings in \cref{Gamma0} that give rise to Feshbach resonances in Li~+~CaH$(v=0,N=0)$ collisions. As a consequence, the density of  magnetic Feshbach resonances does not grow dramatically with adding extra vibrational states to the CC basis set}, as observed in \cref{fig_CSvsB_gs}.  This argument also explains why the resonance density  observed in our previous calculations on spin-polarized Li~+~CaH \cite{Morita:19a}, Li~+~SrOH \cite{Morita:17}, and Rb~+~SrF \cite{Morita:18} collisions is nearly as low as in ultracold atom-atom collisions despite the large number of closed rotational channels.

Significantly, the above argument implies that the computationally expedient rigid-rotor approximation \cite{Tscherbul:11,Morita:17,Morita:18} can be used  to qualitatively predict the spectrum of magnetic Feshbach resonances in ultracold collisions of spin-polarized  $^2\Sigma$ molecules with alkali-metal atoms.  For the Li~+~CaH calculations reported here, the rigid-rotor approximation allows for a 14-fold reduction in the number of scattering channels, decreasing computational cost by over  three orders of  magnitude.

We note that the above conclusion does not necessarily apply to the initial atom-molecule states that are not fully spin-polarized. For example, collisions of CaH ($M_{S_A}=-1/2$) molecules with Li ($M_{S_B}=1/2$) atoms will be affected by the spin-exchange interaction $J(R,r,\theta)\hat{S}_A\cdot \hat{S}_B$, with the prefactor $J(R,r,\theta)$ proportional to the difference between the singlet and triplet PESs in \cref{Hint}. As the spin-exchange interaction depends explicitly on both the vibrational coordinate $r$ and the spin DOF, it will couple the $v=0$ open channel with the other spin states in the $v'>0$ manifolds, potentially leading to significant effects of vibrational DOF on the spectrum of atom-molecule magnetic Feshbach resonances.

\subsection{Vibrational relaxation in Li~+~CaH$(v=1)$ collisions}

Thus far we have considered the dynamics of spin-polarized collisions of Li atoms with CaH molecules in their ground vibrational state.  Ultracold collisions of vibrationally excited CaH molecules with Li atoms are  also of significant interest as they determine the rate of vibrational cooling of trapped molecules \cite{Rellergert:13}. Rapid vibrational energy transfer in atom-molecule collisions would imply efficient sympathetic cooling of vibrational degrees of freedom to produce trapped molecules in their ground vibrational  state.

\begin{figure}[t!]
\begin{center}
\includegraphics[width=\linewidth]{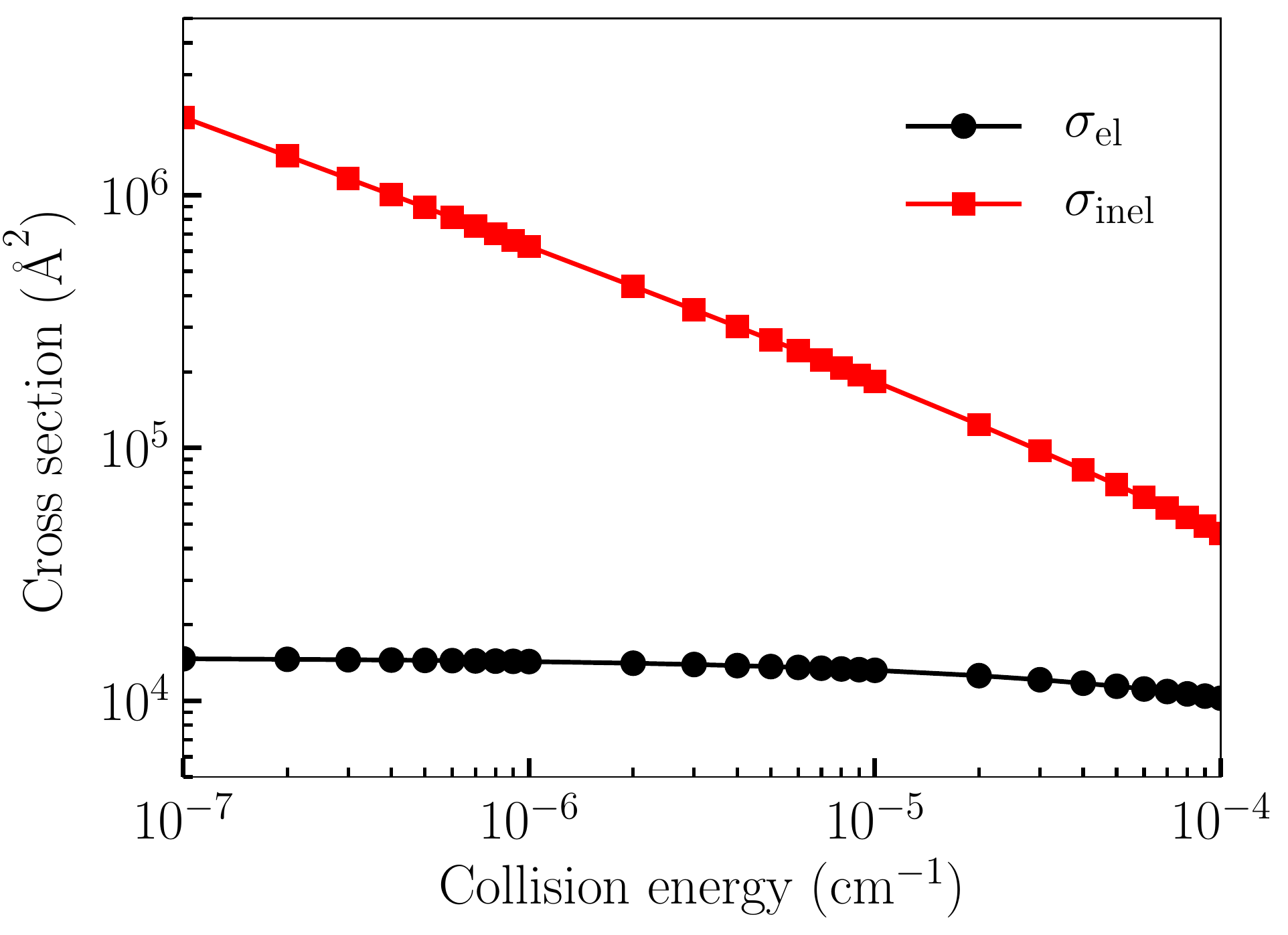}
\end{center}
\caption{Total elastic (black circles) and inelastic (red squares) cross sections for ultracold collisions of vibrationally excited CaH($v=1$, $N=0, M_{S_{A}}=1/2$) with Li($M_{S_{B}}=1/2$)  as a function of collision energy at a magnetic field of 1000~G. 
}
\label{fig_CSvsE_v1}
\end{figure}

In \cref{fig_CSvsE_v1} we compare the total elastic and inelastic cross sections for Li~+~CaH($v=1,N=0$) collisions in a magnetic field of {1000~G}. Inelastic relaxation of vibrationally excited CaH molecules occurs much faster than elastic scattering in the Wigner $s$-wave regime, with $\gamma \ll 1$ at $E_C<0.1$~mK. We verified that the inelastic cross sections for Li~+~CaH($v=1,N=0$)  are insensitive to the magnitude of the external magnetic field between 10 and 2000~G.

This is in striking contrast with ultracold collision dynamics of ground-vibrational-state CaH molecules, which tends to be strongly dominated by elastic scattering ($\gamma\gg 1$) and vary strongly with the field as shown in \cref{fig_CSvsB}. We attribute the large vibrational relaxation rates to the substantial vibrational anisotropy of the Li-CaH interaction PES. As pointed out in Sec. IIA, stretching the CaH vibrational coordinate $r$ results in significant changes in the PES topology, leading to a pronounced $r$ dependence of  the Legendre components $V_\lambda(r,R)$. This enhances the coupling matrix elements $\langle \chi_{vN}|V_\lambda(r,R)|\chi_{v'N'}\rangle$, which drive inelastic transitions between the different vibrational manifolds.

\begin{figure}[t!]
\begin{center}
\includegraphics[width=1.1\linewidth, trim = 60 70 0 50]{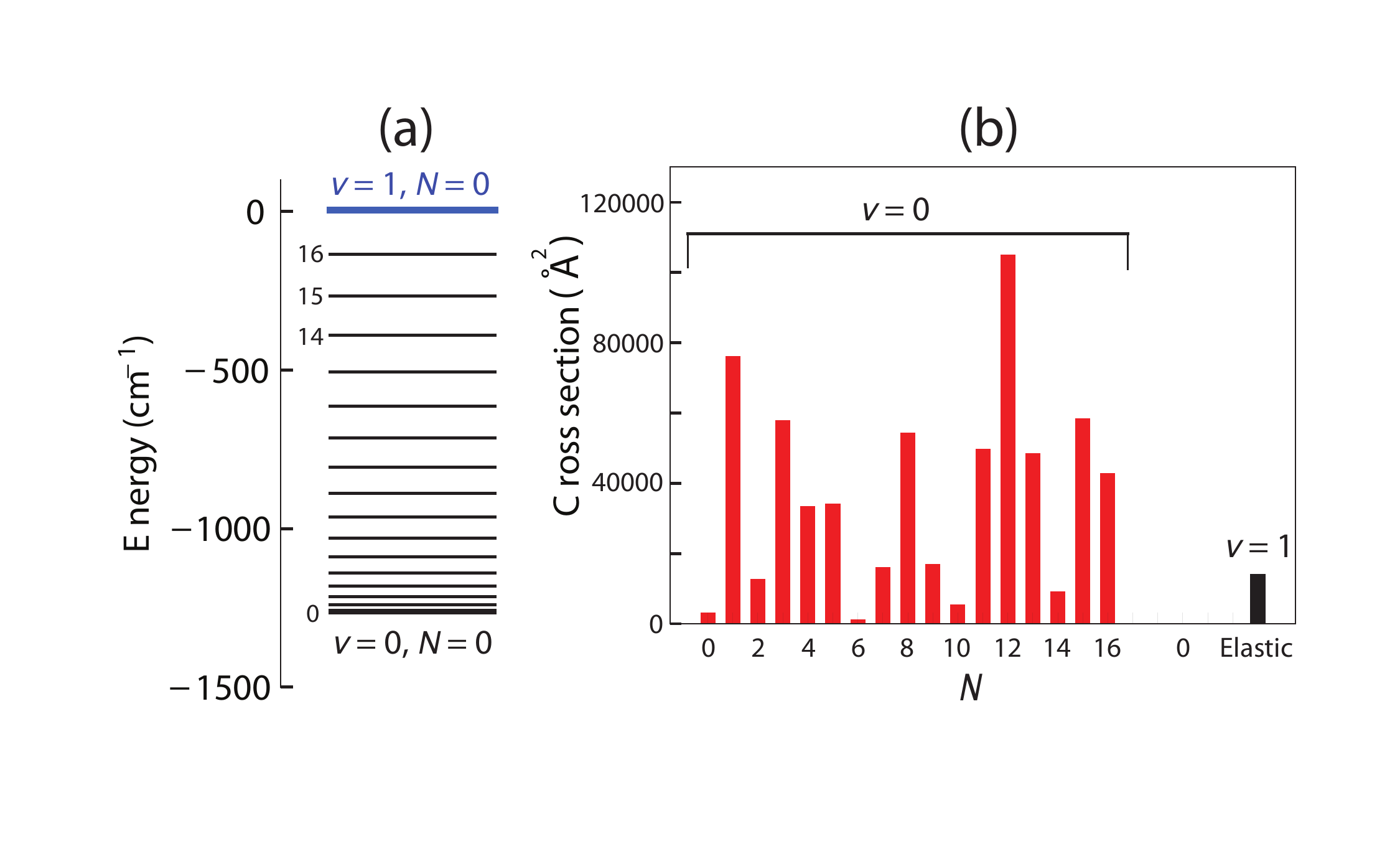}
\end{center}
\caption{(a) Rovibrational energy levels of CaH($^2\Sigma^+$). 
The energies are shown relative to the rovibrational energy of the $v=1$, $N=0$ state at $B=0$. (b) Rotational product state distributions following vibrational quenching of CaH($v=1,N=0,M_{S_{A}}=1/2)$ by Li($M_{S_{B}}=1/2$) at a collision energy of 1.4~$\mu$K and  magnetic field of 1000~G. The following basis set parameters were used in the calculations: $J_\text{max}=1,\, N_\text{max}=55,\, v_\text{max}=13$.
}
\label{fig_rotdist}
\end{figure}

To explore ultracold Li~+~CaH$(v=1)$ scattering in more detail, we examine two key contributions to the total inelastic cross section due to $v$-conserving (``pure'') spin relaxation in the $v=1,N=0$ manifold and the $v=1,N=0\to v=0,N'$ vibrational quenching transitions. 
\Cref{fig_rotdist} shows the inelastic cross sections resolved over final rovibrational states.  We observe that  vibrational quenching accounts for over 99.99\% of the total inelastic cross section. That is, most of the products of collision-induced  vibrational quenching of CaH  $(v=1,N=0,M_{S_{A}}=1/2)$ form in the ground vibrational state.  The vibrationally elastic spin relaxation cross section in the  $v=1$ spin manifold is comparable in magnitude to the $v=0$ spin relaxation cross section (see Sec. IIIB above). The product rotational state distributions  of CaH$(v=0)$ show significant structure, with the population of the $N=0$ and $N=6$ rotational states being strongly suppressed, and that of $N=1$ and $N=12$ strongly enhanced. 
We verified that these results cannot be described by the chaotic Poisson distribution found for the product rotational state distributions of the K~+~KRb chemical reaction \cite{Croft:17}. The lack of chaoticity can be attributed to a small number of product rovibrational states of CaH($v=0$) available due to its large rotational constant (17 compared to 142 for K~+~KRb).

The cross section for $v$-conserving spin relaxation in Li~+~CaH($v=0,N=0$) collisions $\sigma_\text{inel}^{v=1}$ can be obtained by summing the cross sections for the Li$(M_{S_A}=1/2)$~+~CaH$(v=1, N=0,M_{S_B}=1/2)$ $\to$ Li$(M_{S_A}')$~+~CaH$(v=1, N=0,M_{S_B}')$ transitions over all final  $M_{S_A}'\ne M_{S_A}$ and $M_{S_B'}\ne M_{S_B}$. We obtain $\sigma_\text{inel}^{v=1}=  2.41$~\AA$^2$ or 3.85$\times 10^{-4}$\% of the total inelastic cross section, which illustrates that $v$-conserving  spin relaxation, which flips the spin projections without changing the rovibrational state of the molecule,  makes a negligible contribution to the total inelastic cross sections for Li~+~CaH($v=1,N=0$) collisions.
\color{black}

\color{Black}

\subsection{Sensitivity to the interaction potential: Cumulative probability distributions}

In the preceding sections, we used computationally intensive quantum dynamics calculations to calculate converged  cross sections for ultracold scattering Li~+~ CaH in a magnetic field. These calculations are based on a single adiabatic PES, which contains residual inaccuracies due to imperfections of {\it ab initio} methods and fitting errors. 
 As is well known, ultracold scattering observables are very sensitive to these inaccuracies, making quantitative predictions extremely difficult \cite{Morita:19b}. For example, a 1\% change in the PES depth, which is smaller than the typical uncertainty of state-of-the art {\it ab initio} interaction potentials (such as the one developed in Sec.~II), can modify the elastic and inelastic cross sections by many  orders  of magnitude due to the presence of numerous scattering resonances \cite{Suleimanov:16,Morita:19b}.

We have recently suggested a probabilistic approach to this problem based on cumulative probability distributions (CPDs) obtained by averaging scattering observables  results over an ensemble of slightly different interaction PESs \cite{Morita:19b}. 
Specifically, the CPD of a scattering observable $\gamma$ (such as the elastic or inelastic cross section or their ratio) is defined as \cite{Morita:19b}
\begin{equation}\label{eq:CP}
F(\Gamma) = P(\gamma\le\Gamma) = \frac{N(\gamma\le\Gamma)}{N_\text{tot}},
\end{equation}
where $P(\gamma\le\Gamma)$ is the probability that $\gamma$ is less than $\Gamma$. In the limit of infinitely large ensemble size, the CPD may be expressed as $F(\Gamma) =  \frac{N(\gamma\le\Gamma)}{N_\text{tot}}$, where $N(\gamma\le\Gamma)$ is the number of samples, for which $\gamma<\Gamma$ and $N_\text{tot}$ is the ensemble size.   Alternatively, $F(\Gamma) = \int_{0}^{\Gamma} p(\gamma)d\gamma$, where $p(\gamma)$ is the probability density function (PDF) of the observable $\gamma$. Given the  PDF, the probability for $\gamma$ to fall within a specific range of values $P(a<\gamma<b)=\int_a^b p(x)dx$.
For example, the probability that a collision system has a ratio of elastic to inelastic cross sections in excess of 100 (a favorable condition for sympathetic cooling) is given by  $P_s=1-F(\Gamma=100)$. Thus, the CPDs could be used to  screen atomic and molecular candidates with favorable properties for sympathetic cooling experiments \cite{Morita:19b}.
Significantly, because the CPDs are much less affected by narrow scattering resonances than single-PES results, they do not require complete basis set convergence, and hence can be computed much more efficiently by using severely restricted coupled-channel basis sets~\cite{Morita:19b}.

Our calculations used a sample of $N_\text{tot}=201$ interaction PESs $V_i$ obtained by scaling the original interaction PES (see Sec. II) by a constant factor $\lambda\in  0.95{-}1.05$ as described in Ref.~\cite{Morita:19b}. This corresponds to a 5\% uncertainty in our \textit{ab initio} Li-CaH PESs, which is the typical margin of error of modern high-level {\it ab initio} methods for triatomic atom-molecule collision complexes \cite{Morita:19b}.
By carrying out CC calculations for every PESs $V_i$ in the ensemble $\{V_i$\}, we obtain samples of  elastic $\left\{  {\sigma_{\mathrm{el},i}} \right\}$ and inelastic $\left\{  {\sigma_{\mathrm{inel},i}} \right\}$ cross sections and of their ratio $\left\{\gamma_{i}  \right\}$. Finally, we evaluate the number of samples $N(\gamma\le\Gamma)$, for which the relation $\gamma\le\Gamma$ is satisfied as a function of $\Gamma$, and calculate the CPD using Eq. (\ref{eq:CP}).

\begin{figure}[t!]
\begin{center}
\includegraphics[width=\linewidth]{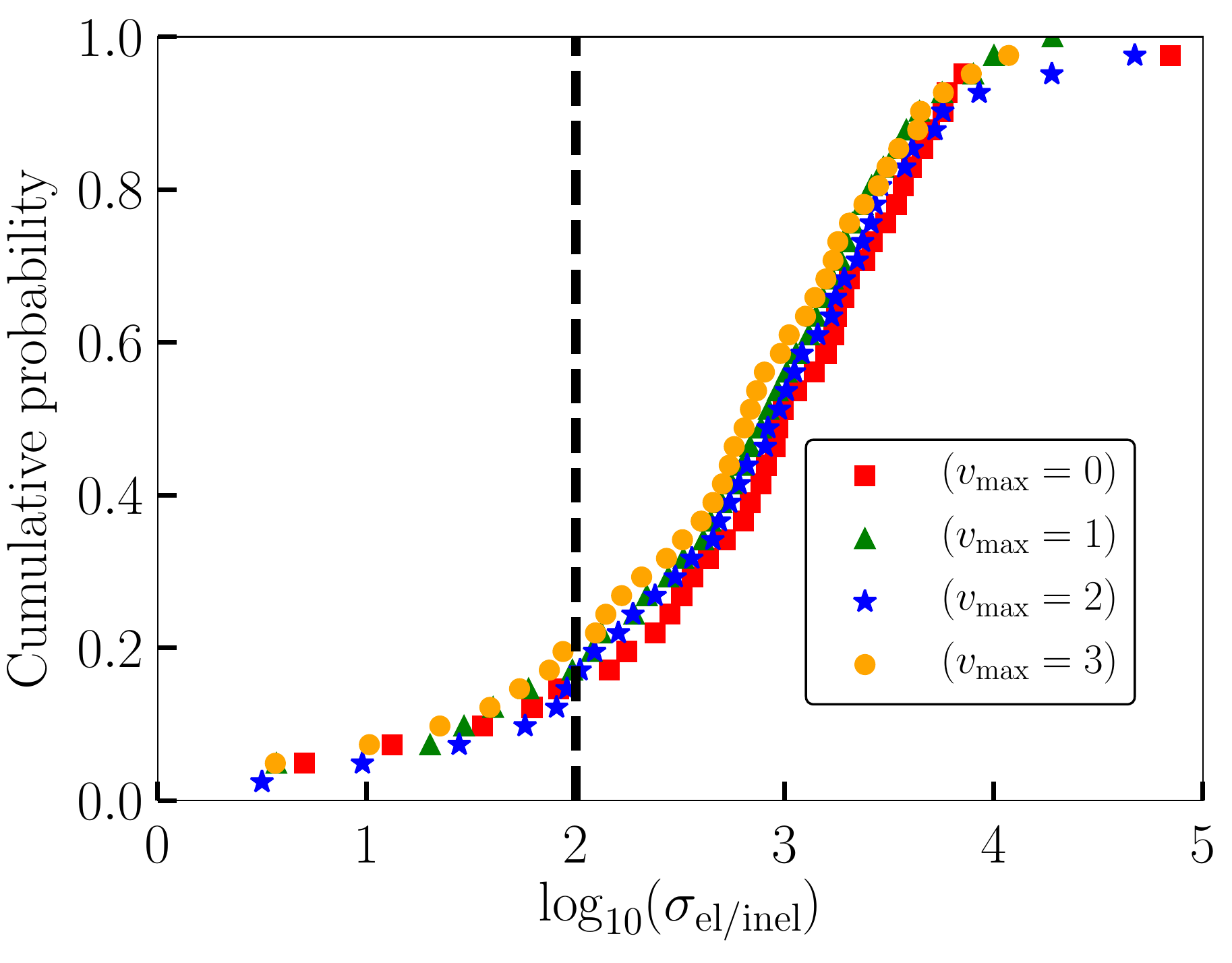}
\end{center}
\caption{
Cumulative probability distributions of the elastic-to-inelastic ratio $\gamma=\sigma_\mathrm{el}/\sigma_\mathrm{inel}$ for spin-polarized Li~+~CaH ($v=0$, $N=0$) collisions calculated with $v_\text{max}=0$ (red crosses), $v_\text{max}=1$ (green triangles), $v_\text{max}=2$ (blue stars) and $v_\text{max}=3$ (orange circles).
The averaging is performed over an ensemble of  201 interaction PESs obtained by multiplying the original 3D PES (Sec. IIA) by a constant factor $\lambda = 0.95,...,1.05$.  The collision energy is $10^{-4}$ cm$^{-1}$ and the magnetic field is 1000 G. 
}
\label{fig_Cumulative_vmax_CS}
\end{figure}

To explore the effect of vibrational DOF on the CPDs, we plot in \cref{fig_Cumulative_vmax_CS}  the probability distributions of the ratio of the elastic to inelastic cross sections for Li~+~CaH $P(\gamma)$.
The distributions obtained with one, two, and three vibrational states in the basis set are similar to each other and to the $v_\text{max}=0$ CPD.  
This can be explained by analyzing the $\lambda$ dependence of the cross sections \cite{Morita:19b}. While adding vibrational DOF increases the number of resonances in the cross sections as a function of $\lambda$, these resonances are averaged out when integrated over a sufficiently large interval of $\lambda$.
We note that the differences between the CPDs for different $v_\text{max}$ shown in \cref{fig_Cumulative_vmax_CS} cannot be regarded as physically significant due to   incomplete convergence of scattering observables and a limited PES sample size, which limit the accuracy of the calculated CPDs to 15-20\%.

The resulting distributions all predict a large cumulative probability $P (\gamma\ge 100) \simeq 80\%$, which is  favorable for atom-molecule sympathetic cooling \cite{Tscherbul:11}. The same conclusion was reached in our previous work using the rigid-rotor approximation \cite{Morita:19b}, suggesting that  the approximation can be applied to describe  the statistical  properties of ultracold Li~+~CaH($v=0,N=0$) collisions.

%

\section{Summary}

We have presented a computationally efficient methodology, based on the total angular momentum representation \cite{Tscherbul:10}, for full-dimensional quantum scattering calculations on atom-molecule collisions in the presence of an external magnetic field. We apply the methodology to elucidate the magnetic field dependence of scattering cross sections for strongly anisotropic collisions of  $^2\Sigma$ molecular radicals (CaH) with ultracold $^2$S atoms (Li).
To achieve numerical convergence of scattering observables, we used an extended CC basis set including 14 vibrational and 56 rotational states of CaH (per each vibrational state), along with the spin states of the collision partners. We developed a new {\it ab initio} 3D PES for the triplet state of the  Li-CaH collision complex featuring an explicit dependence on the CaH internuclear distance $r$ using a highly correlated UCCSD[T] method and a large basis set. The PES is 7202.4 cm$^{-1}$ deep and it is also  strongly anisotropic.

Our full-dimensional quantum  scattering calculations are in qualitative agreement with the previous reduced-dimensional Li~+~CaH   calculations using the rigid-rotor approximation \cite{Tscherbul:11}. Both calculations predict large ratios of elastic to inelastic cross sections  ($\sigma_\text{el}/\sigma_\text{inel}>100$), indicating favorable prospects for sympathetic cooling of trapped CaH molecules with Li atoms.
This indicates  that the rigid-rotor approximation can provide a  qualitatively accurate description of ultracold spin-polarized collisions of  $^2$S atoms with open-shell $\Sigma$-state molecules in their ground vibrational state.
However,  a quantitatively accurate description of the inelastic cross sections as a function of external magnetic field requires the use of a complete rovibrational basis set including 14 vibrational states.

The full-dimensional treatment of the atom-molecule collision problem allowed us to explore vibrational relaxation of CaH$(v=1,N=0,M_S=1/2)$ molecules in ultracold collisions with Li atoms. 
We find that vibrational quenching is extremely fast, suggesting that vibrational cooling of CaH molecules by Li atoms will be efficient.  The quenching cross section is independent of the magnitude of the external magnetic field.

Our calculations uncover the role of vibrational DOF on the density of magnetic Feshbach resonances in ultracold atom-molecule collisions, an important quantity that determines the extent of  quantum chaos, complex formation, and sticking in ultracold molecular collisions \cite{Mayle:12,Mayle:13,Frisch:14}. We find that adding a large number of closed rovibrational channels, while important to achieve numerical convergence, does not significantly alter the mean density of magnetic Feshbach resonances in collisions of spin-polarized atoms and molecules in their absolute ground states. We suggest a tentative explanation for this surprising result: Because magnetic (spin-dependent) couplings in the atom-molecule Hamiltonian are largely independent of $r$, the internuclear distance of the diatomic molecule,  closed  rovibrational channels with $v' \ge 1$ remain uncoupled from the $v=0$ incident open channel, dramatically reducing the closed-channel subspace available for the formation of magnetic Feshbach resonances. To rigorously test this hypothesis,  it would be desirable to calculate the near-threshold bound states of the collision complex as a function of external magnetic field and locate their crossings with atom-molecule  thresholds, which correspond to zero-energy Feshbach resonances \cite{Chin:10}.

In future work, we plan to explore ultracold collisions of atoms and molecules in non fully spin-polarized initial states. This would require explicit treatment of the low-spin atom-molecule PES of singlet symmetry, which could be realistically accomplished for non-reactive collision systems such as  Rb~+~SrF  \cite{Kosicki:17} using the total angular momentum representation \cite{Tscherbul:10,Tscherbul:11,Morita:17,Morita:18}.
Such a study could reveal significant vibrational effects on the atom-molecule resonance density due to the  $r$ dependence of the Heisenberg exchange interaction. 
 It would also be interesting to explore whether the $r$ dependence of the other spin-dependent interactions (such as the intramolecular spin-rotation interaction in CaH and/or the intermolecular spin-orbit interaction in Li-CaH \cite{Warehime:15})  could induce magnetic Feshbach resonances in ultracold atom-molecule collisions.

\begin{acknowledgments}
We are grateful to Dr. Ivan Kozyryev for stimulating discussions. This work was supported by NSF Grants No. PHY-1607610 and PHY-1912668, and partially by the U.S. Air Force Office for Scientific Research (AFOSR) under Contract No. FA9550-19-1-0312. J. K. acknowledges financial support from the U. S. National Science Foundation, Grant No. CHE-1565872 to Millard Alexander. J.~K. acknowledges XSEDE.org Grant
No.\ CHE-130120 and University of Maryland High-Performance Deepthought facilities for computational time.
\end{acknowledgments}
\newpage

\section*{Appendix}

\subsection{Rotational basis set convergence}

\begin{figure}[hbt!]
\begin{center}
\includegraphics[width=\linewidth]{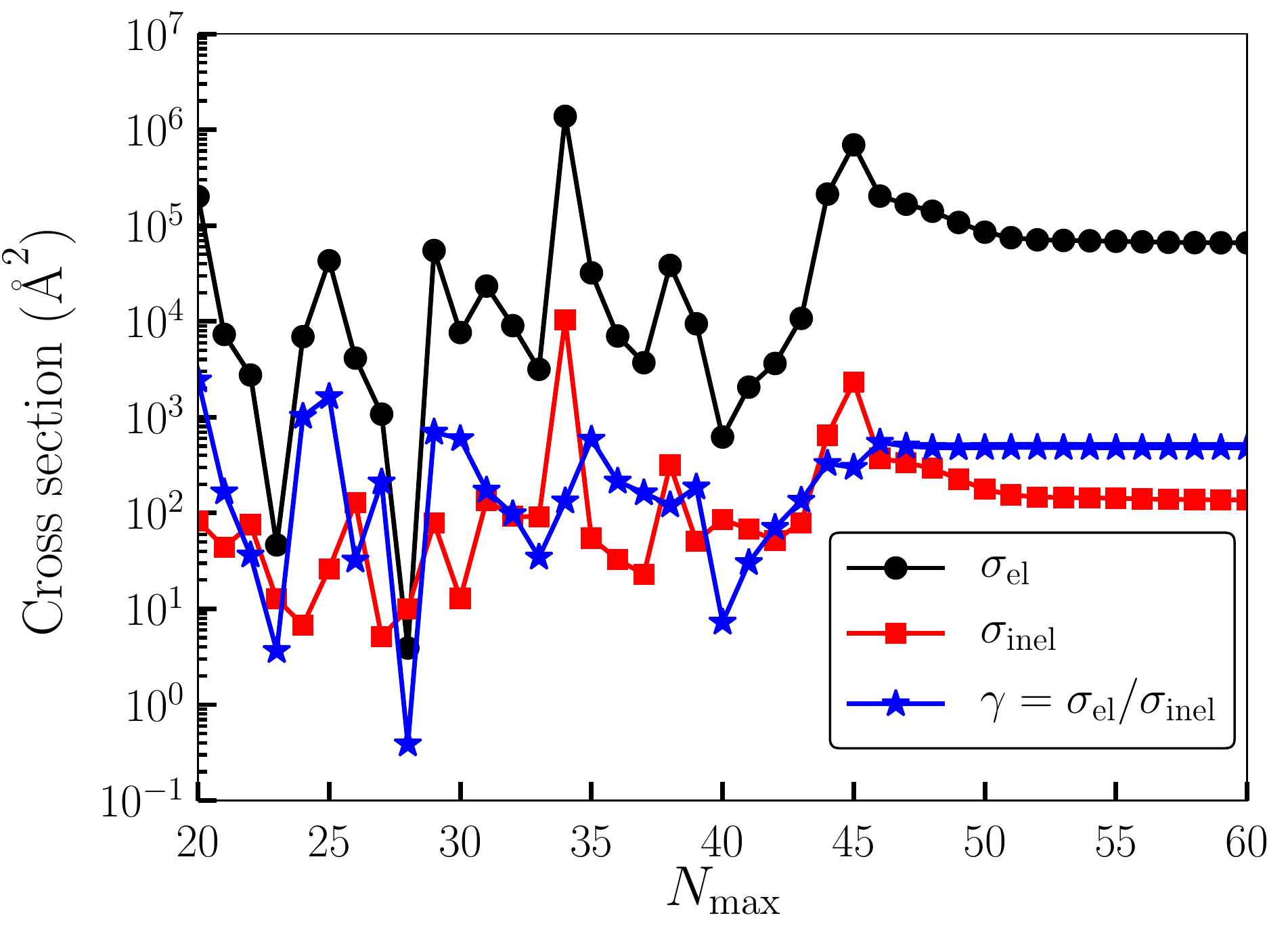}
\end{center}
\caption{Elastic cross sections (black circles), inelastic cross sections (red squares), and the elastic-to-inelastic ratios (blue stars) for Li~+~CaH as a function of the highest rotational state of CaH  in the basis. The collision energy $E_C=10^{-6}$ cm$^{-1}$, magnetic field $B= 1000$~G, $v_\text{max}=0$, and $J_\text{max} = 1$.
}
\label{fig_Conv_Nmax}
\end{figure}

 To study rotational basis set convergence, we plot in  \cref{fig_Conv_Nmax} the calculated spin relaxation cross section as a function of $N_\text{max}$, the number of CaH rotational states in the CC basis set  containing  a single $v=0$ vibrational state.
 As discussed previously \cite{Tscherbul:11}, the large anisotropy and depth of the Li-CaH interaction couple  a large number of rotational states of CaH. As a result, we observe rapid oscillations of scattering cross sections as a function of  $N_\text{max}$ up to $N_\text{max}\leq 45$. A monotonic change of the cross sections continues until $N_\text{max}\sim50$. Interestingly, the elastic-to-inelastic ratio exhibits slightly faster convergence at around $N_\text{max}=45$, likely due to correlated  residual convergence errors.
 The requirement of large $N_\text{max}>50$ to obtain convergence is consistent with the results of the previous studies using the rigid-rotor approximation \cite{Tscherbul:11,Morita:18,Morita:19b}. 
In the calculations reported in the main text, we employ $N_\text{max}=55$ unless otherwise stated.

\subsection{Effect of PES scaling}

\begin{figure}[t!]
\begin{center}
\includegraphics[width=\linewidth]{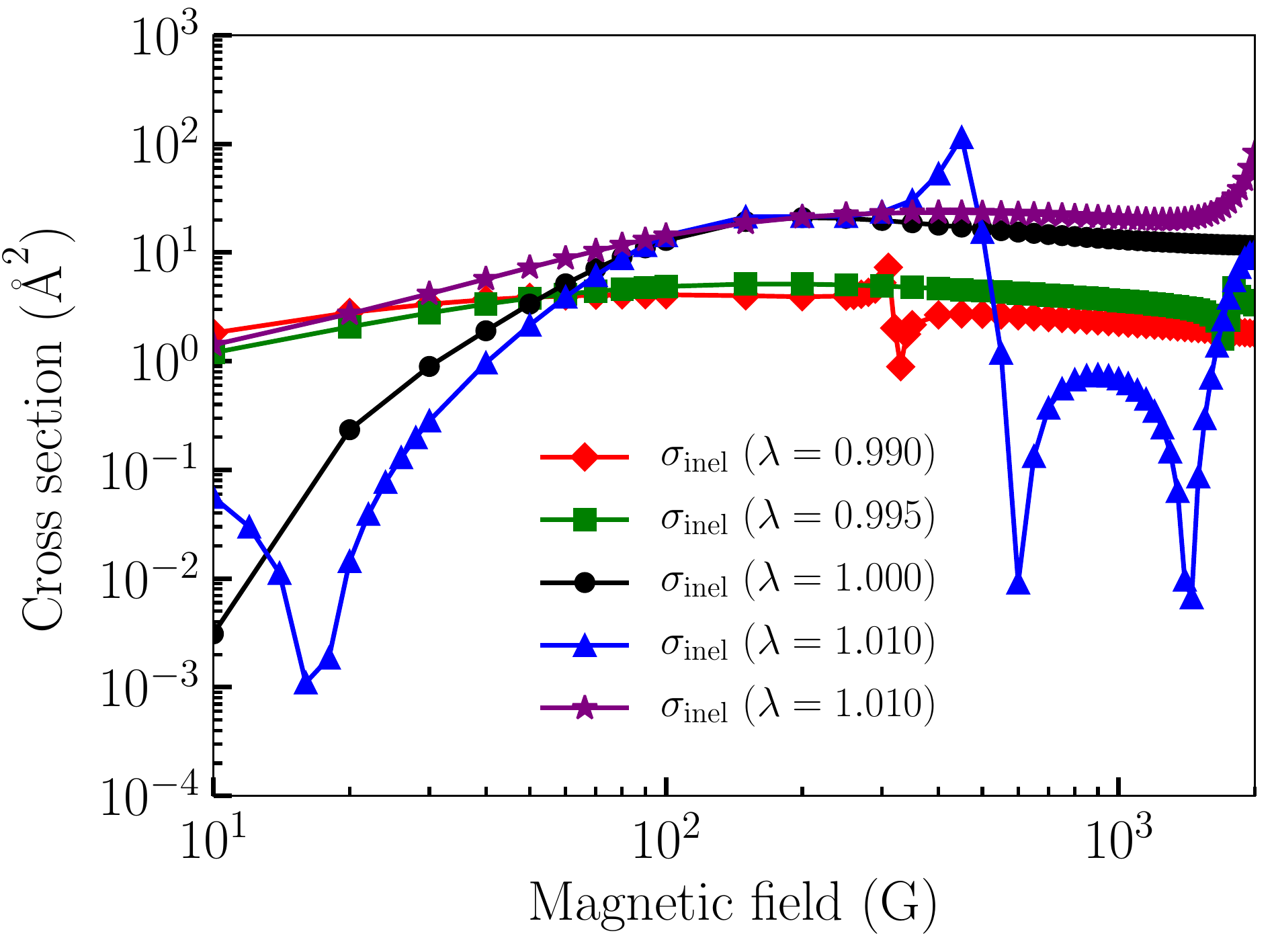}
\end{center}
\caption{Magnetic field dependence of inelastic cross sections for spin-polarized Li~+~CaH ($v=0$, $N=0$) collisions at the collision energy of $10^{-4}$ cm$^{-1}$ with the interaction potentials scaled by a factor of $\lambda=0.990$ (red crosses), $\lambda=0.995$ (green squares), $\lambda=1.000$ (black circles),  $\lambda=1.005$ (blue triangles) and $\lambda=1.010$ (purple stars). 
$v_\text{max}=0$ and $N_\text{max} = 55$.
}
\label{fig_lambda_v0_CSvsB_1e-4cm}
\end{figure}

To explore whether the 30~G resonance in the fully converged 3D results shown in \cref{fig_CSvsB} can be reproduced in truncated  $v_\mathrm{max}=0$ calculations with a scaled interaction PES, we plot in \cref{fig_lambda_v0_CSvsB_1e-4cm} the spin relaxation cross sections calculated with several different interaction potentials obtained by multiplying the original PES by a constant scaling factor $\lambda$.  We see that while the magnetic field dependence calculated for $\lambda=1.01$ shows a peak near 50~G, the asymmetric shape of the peak is completely different from that of the 30~G resonance shown in \cref{fig_CSvsB}. Thus, $\lambda$ scaling  of the interaction PES cannot reproduce the resonance features seen in full-dimensional calculations.

%

\end{document}